# First GNSS-deployed optical clock for local time scale upgrade

Yi Yuan[1,†], Jian Cao[1,†,*], Jinbo Yuan[1,†], Dehao Wang[1], Pengcheng Fang[1], Qunfeng Chen[1], Shiying Cao[2], Xuanjian Wang[1], Sijia Chao[1], Hualin Shu[1], Guojun Li[3,*], Jinfeng Xu[3], Guitao Fu[3], Yuting Yang[3], Run Zhao[3], Fengfeng Shi[3], Xueren Huang[1,4,*]

[1] Key Laboratory of Time Reference and Applications, Innovation Academy for Precision Measurement Science and Technology, Chinese Academy of Sciences, Wuhan 430071, China
[2] National Institute of Metrology, Beijing 100049, China
[3] Beijing Satellite Navigation Center, Beijing 100000, China
[4] Wuhan Institute of Quantum Technology, Wuhan 430206, China
† These authors contributed equally to this work
* E-mail:caojian@apm.ac.cn, 1010551750@qq.com, hxueren@apm.ac.cn.

**Abstract**

Precise time scale is the universal base for all measurements. Here we report the deployment of a compact and transportable optical clock to a timekeeping institution and steering an active hydrogen maser to generate an optical time scale, realizing the upgrade of the local time scale in the Global Navigation Satellite System. The optical clock was transported over 1200 km by express delivery and resume work as normal promptly, and its extremely high uptime of 93.6% in the half-year enabled us to precisely correct the frequency drift of hydrogen maser, ultimately achieving an unprecedented monthly instability of $4\times10^{-17}$. This steering experiment with a deployable optical clock marks a significant advancement, demonstrating that a timing accuracy below 100 ps per month can be achieved feasibly in various timekeeping institutions where hydrogen masers are typically employed as the primary contributor to timekeeping. In the future, mobile optical time scale based on such transportable optical clock can be deployed flexibly and rapidly, which is particularly important in scenarios lacking International Atomic Time reference.

## 1. Introduction

Atomic time scale constitutes the backbone of modern sciences and technologies, with applications ranging from fundamental physics to Global Navigation Satellite System (GNSS) for positioning and synchronization. International Atomic Time (TAI), as a resultant time scale with the accuracy and monthly instability of $10^{-16}$ [1,2], is synthesized from ~500 microwave atomic clocks worldwide due to their superior reliability. Limited by the frequency stability of these clocks and the accuracy of primary frequency standards based on Cs fountain clocks, so the accruing timing error of TAI is still at the level of nanosecond [1,2].

The performance of optical clocks is two orders of magnitude more accurate and stable than their microwave counterparts [3–7], which provides a blueprint for establishing an Optical time scale (OTS) to upgrade the ongoing TAI [8–15]. The prerequisites for realizing this blueprint mainly contain deployable optical clocks and their seamless connection to existing hardwares of TAI. Compact and transportable optical clock (TOC) with high reliability and acquirebility [16–22], which means the ability to be flexibly deployed worldwide, is well suited for this application scenario. Moreover, the contribution of active hydrogen masers (HMs) to the formation of TAI is greater than 90% because they have almost the best frequency stability and reliability of commercial clocks [23]. However, the frequency of HMs is subject to a typical long-term drift of $10^{-16}$/day, which makes them strongly dependent on the correction of better references during the national time services. Therefore, using TOCs with frequency instability and uncertainty at $1\times10^{-17}$ or below and steering HMs to the level of $10^{-17}$ is expected to pave the way for upgrade of TAI to sub-nanosecond levels. In this study, we successfully enhanced a local time scale in the Global Navigation Satellite System by steering an active hydrogen maser with a transportable optical clock, achieving unprecedented stability ($4\times10^{-17}$@1 month) and sub-100-picosecond timing accuracy. This demonstrates the feasibility of deployable optical clocks for precision timekeeping applications.

## 2. Deployment of the Optical Clock

Building on our previous advancements in TOCs [16,19,24,25], we have developed a transportable $^{40}Ca^+$ optical clock (TOC-729-3) with a compact total volume of ~1.5 m³ with an operating power of ~600 W. To optimize its portability, compactness and robustness, the transportable $^{40}Ca^+$ optical clock is designed with a modular architecture, comprising three independent cabinets. These modules are engineered to be interchangeable across our multiple optical clock systems. As illustrated in the inset of Figure. 1A, the two racks on the outside (left and right) house the electronic control units, which integrate critical instruments such as timing sequence controllers, external-cavity diode lasers (ECDLs), and signal amplifiers. The central rack, dedicated to the physical system of clock, is divided into three functional sections with a total volume of ~0.4 m³, as detailed in Figure. 1B. The bottom section contains a multi-channel ultra-stable cavity for all lasers to suppress long-term frequency drift, among which the short-term frequency instability of the 729 nm clock laser is better than $2\times10^{-15}$ with an averaging time between $1\sim20$ s [26]. The middle section features two retractable drawers that store laser distribution components, including devices for laser frequency shifting, beam splitting, and fiber noise cancellation. The 729 nm laser is split into four beams, which are precisely directed to the wavelength meter, reference cavity, ion trap, and optical frequency comb (OFC), respectively. Fiber noise is cancelled for all outputs except the beam directed to the wavelength meter. The top section accommodates the linear ion trap and fluorescence collection system, which are essential for ion confinement and state detection. Compared to the previous iteration of TOCs, the newly developed clock incorporates advanced automated programs that significantly enhance operational uptime and system reliability. And the systematic evaluation confirms that the total frequency shift uncertainty is at the level of $1\sim2\times10^{-17}$ (see Supplementary material).

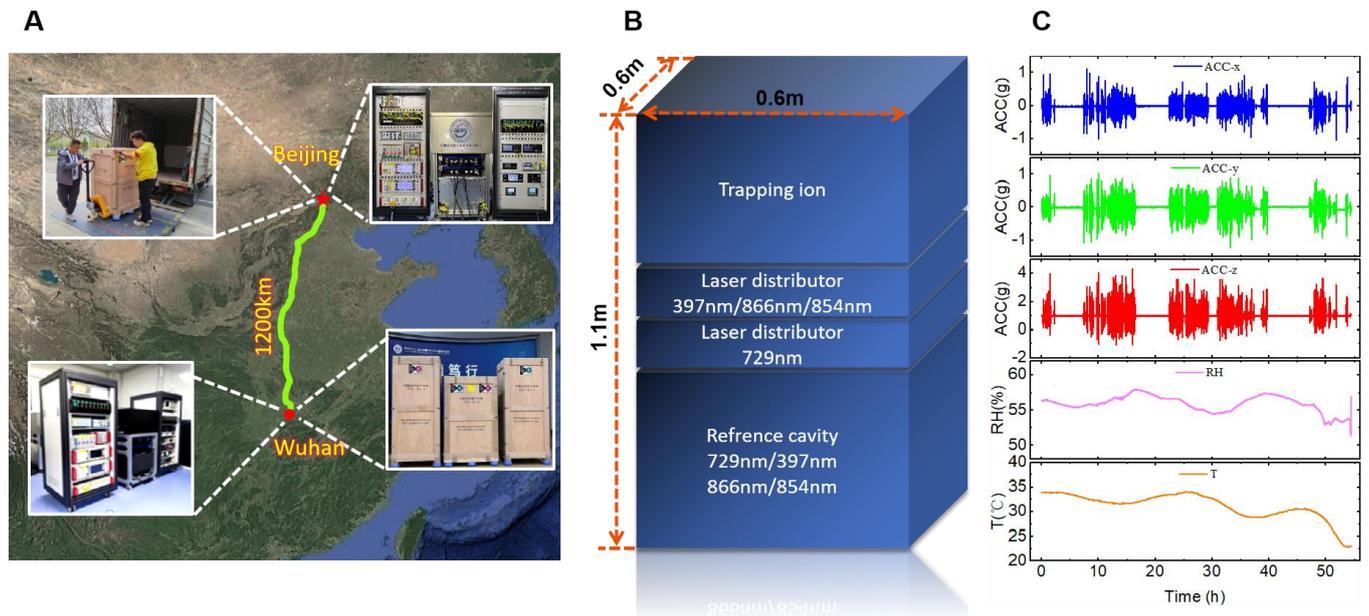

**Figure 1. Transportation of $^{40}Ca^+$ optical clock (TOC-729-3).** (**A**) The transportation route from Wuhan to Beijing. The insets show the appearance of TOC-729-3 in Wuhan, in Beijing and during the transportation. (**B**) The design scheme of the physical system of TOC-729-3. It's composed of four parts, including a trapping ion box, two laser distributor boxes, a reference cavity box, and all of them are integrated in a 0.6 m×0.6 m×1.1 m rack. (**C**) Records of the environmental parameters during transportation of the TOC-729-3. The blue, green, red lines represent the acceleration in three directions with a sample interval of 5 ms, the pink and orange lines represent the temperature and humidity with a sample interval of 1 s.

As illustrated in Figure. 1A, the clock was transported via truck from Wuhan to Beijing and subsequently installed at Beijing Satellite Navigation Center (BSNC), covering a distance exceeding 1200 km without external power supply.

During this transportation, various environmental parameters including vibration, temperature and humidity were meticulously monitored with senors installed in the physical system. As illustrated in Figure. 1C, the horizontal acceleration remained within a range of approximately ±1 g (where 1 g ≈ 9.8 m/s$^2$), while the vertical acceleration varied between -1 g and +4 g. Notably, the humidity fluctuated by more than 5%, and the temperature fluctuated by over 10 °C. Following transportation, the single ion was successfully restored within a single day. Part of this time was spent on the connection of various optical fibers and cables, and the other part was used to optimize the operating parameters of ECDLs. Importantly, no significant deviations were observed in the system components. Thanks to the robust and reliable design of the system, the clock achieved closed-loop locking within two days and was activated to steer the frequency of local HM within two weeks, during which most of the time was cost in temperature recovery of the ultra-stable cavity due to its ultra-long thermal time constant. This achievement vividly demonstrates its outstanding adaptability and stability throughout the transportation and reinstallation process, highlighting its resilience in maintaining optimal performance under such circumstances, and showcasing its excellent robustness and rapid deployment capabilities.

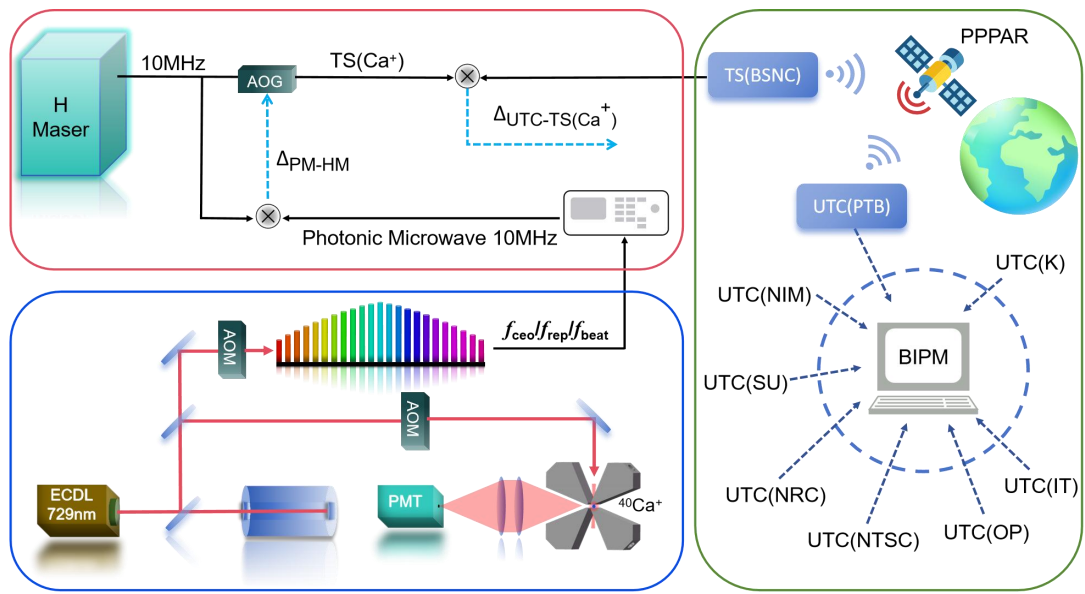

**Figure 2. Schematic diagram of the steering experiment and evaluation of the optical time scale.** Blue block diagram: The $^{40}Ca^+$ optical clock outputs a narrow linewidth laser referenced on 729 nm clock transition to the OFC and related microwave signals of $f_{rep}$/$f_{ceo}$/$f_{beat}$ are transferred to the DCE. Red block diagram: The DCE transforms the input signals into a 10 MHz microwave for comparison with the 10 MHz microwave sourced from the local HM. The difference between them is computed via the steering algorithm, and the output signal of AOG is corrected to produce a real-time time scale TS(Ca$^+$). Subsequently, TS(Ca$^+$) is compared with TS(BSNC) to assess its performance. Green block diagram: TS(BSNC) is compared with UTC(PTB) through a satellite link to finally determine the difference between TS(Ca$^+$) and UTC.

## 3. Experimental Scheme

The experimental platform for generating a high-performance atomic time scale comprises three fundamental parts: (1) a $^{40}Ca^+$ optical clock, which establishes the absolute reference for high-precision frequency measurement; (2) an OFC that acts as a flywheel oscillator, facilitating the optical-to-microwave frequency conversion; and (3) a continuously operating, calibrated timekeeping system that integrates and maintains the temporal reference. This configuration ensures the stability and accuracy required for advanced time scale generation. Once the closed-loop locking and systematic shifts correction of the optical clock are accomplished, the output of 729 nm laser will maintain almost the same frequency accuracy and stability as the clock transition of the $^{40}Ca^+$ ion. Before the steering experiment, the 729 nm laser is first sent to an Er doped fiber femtosecond OFC [27,28] to obtain the beat frequency $f_{beat}$. And then, it is fed into the optical-to-microwave down conversion electronics (DCE) with the carrier-envelope offset frequency $f_{ceo}$ and the repetition frequency $f_{rep}$ from the OFC. The DCE operates based on the principle of transfer oscillator and generates a photonic microwave (PM) of 10 MHz (see Supplementary material) [29,30], which is the critical signal for the steering experiment. The scheme and apparatus of the steering experiment are shown in Figure. 2.

The frequency offset and drift rate of the local HM are obtained through comparing its 10 MHz output signal with the PM of 10 MHz generated by DCE. Subsequently, the correction value for the local HM is calculated through the steering algorithm (see Supplementary material), and then this correction value is used by the auxiliary output generator (AOG) to correct the output of the HM, thereby generating the precise time scale TS(Ca$^+$) referenced to the $^{40}$Ca$^+$ optical clock. Through the satellite link and in combination with the precise point positioning ambiguity resolution (PPPAR) method [31,32], The time scale TS(BSNC) generated by the BSNC is continuously compared with UTC(PTB), thereby determining the time difference between TS(BSNC) and UTC. Taking TS(BSNC) as a bridge, TS(Ca$^+$) is compared with it and ultimately the time difference between UTC and TS(Ca$^+$) is obtained, which is employed to evaluate the performance of the time scale generated by steering system.

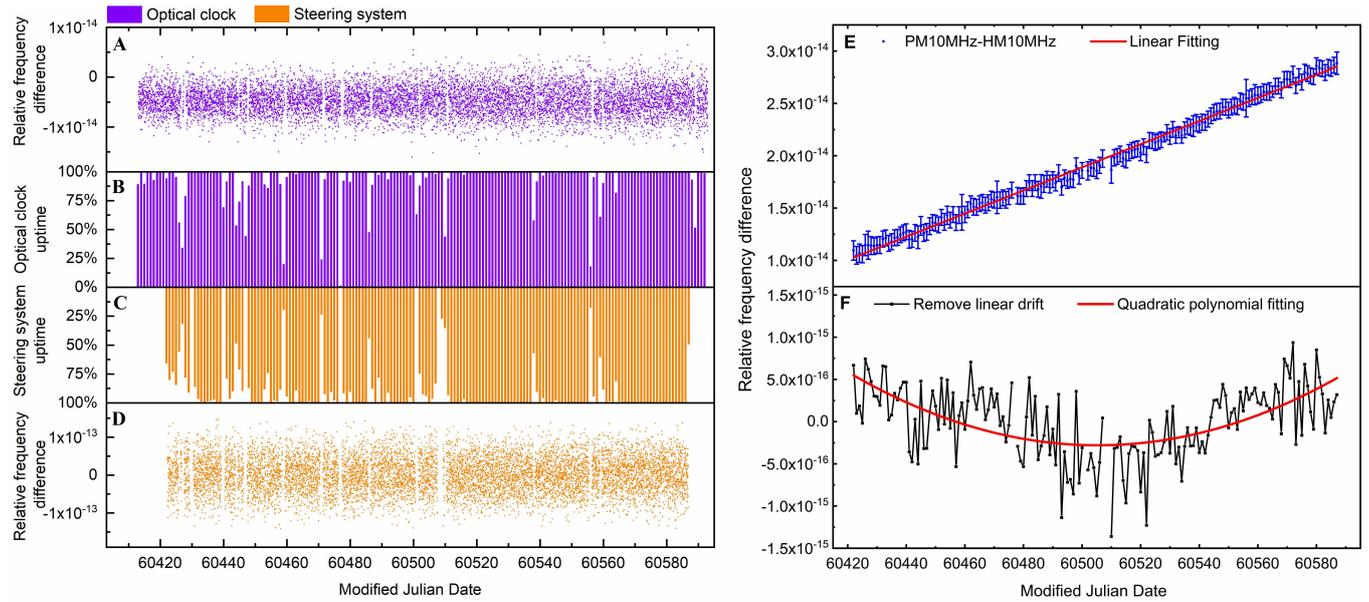

**Figure 3. The uptime of the steering system and the results of relative frequency difference.** (A) The data records of the relative frequency difference between the Δm=0 and Δm=±1 transitions in the $^{40}$Ca$^+$ optical clock. Due to the characteristics of the electric quadrupole transitions in the $^{40}$Ca$^+$ optical clock, the mean relative frequency difference of the two pairs of spectrum is nonzero. The output laser of the optical clock linked to the OFC is referenced to the averaging frequency of the three pairs of spectrum with Δm = 0, ±1, ±2, and the net electric quadrupole frequency shift can be cancelled to zero. (B) The uptime of the $^{40}$Ca$^+$ optical clock during the half a year. (C) The uptime of the steering system during the steering experiment. It is the intersection of the normal operation times of all instruments. (D) and (E), The data records of the relative frequency difference between the local HM and the PM. The difference between them is averaged with the original data from 0:00 to 24:00 (UTC+0) every day into one point. (F) The nonlinear relative frequency drift of the HM. It's obtained by removing the linear drift and the offset of the relative frequency difference. The red solid line represents the quadratic polynomial fitting result of the nonlinear frequency drift of the HM.

## 4. Experimental Results

The experiment of steering the local HM with the $^{40}$Ca$^+$ clock lasted for about half a year. As illustrated in Figure. 3B, the $^{40}$Ca$^+$ clock exhibited an operational uptime of 93.6% for 180 consecutive days from Modified Julian Date (MJD) 60413 to 60592, and the corresponding original data is presented in Figure. 3A. The entire steering system was in operation from MJD 60422 to 60587, and its uptime was mainly limited by the optical clock and the OFC. During this period, due to the uptime of the OFC being approximately 94.5%, the uptime of the steering system was 88.5%, as shown in Figure. 3C. Notably, in the first half of this experiment, the uptime was slightly lower than that in the second half due to factors such as personnel activities the room. This issue was improved after reducing these activities in the second half of period. During the 166-day steering experiment, the relative fractional drift rate of the local HM was measured to be 1.11×10$^{-16}$ per day, and the fitting result is shown as a red solid line in Figure. 3E. Finally, the residual nonlinear drift after removing the linear drift of the HM is presented in Figure. 3F, and it can be seen that its fluctuation range exceeded 1×10$^{-15}$. Furthermore, there are obvious higher-order components in this drift, while optical clocks can precisely measure these nonlinear characteristics and implement the correction.

Before the steering experiment, the noise floor of the DCE was also measured by comparing the output from two sets of independent DCEs referred to the same OFC. The frequency instability contributed by DCE is shown by the purple dotted line in Figure 4A, and the frequency instability of the $^{40}$Ca$^+$ optical clock is also represented by the black solid line in the same figure. It can be seen that an average time of 30000 s is required for the influence of

DCE to be less than that of the optical clock.

During the steering experiment, the 10 MHz signal $f_{HM}$ of the local HM was compared with the 10 MHz signal $f_{PM}$ of the DCE via a phase comparator, which determined the frequency deviation $\Delta f$ between $f_{HM}$ and $f_{PM}$. The instability of $\Delta f$ is indicated by the blue solid line in Figure. 4A, reaching a minimum value of approximately $3.0\times10^{-16}$ around $10^5$ s. Consequently, we performed a correction on the frequency drift of the HM once a day. By statistically analyzing the data set of $\Delta f$ from 0:00 to 24:00 (UTC+0) of the previous day, the correction value $f_{steer}$ for the current day was obtained (see Supplementary material). The red solid line in Figure. 4A shows the frequency instability of $f_{HM}$' after the HM is corrected relative to $f_{PM}$, which can be approximately regarded as the frequency instability of TS(Ca$^+$) relative to the $^{40}$Ca$^+$ optical clock. This instability continuously improves after $10^5$ s, thus achieving the goal of steering the HM using the optical clock. When the averaging time is one month, the instability corresponding to the blue solid line reaches $3.0\times10^{-15}$, and the instability corresponding to the red solid line reaches $4.2\times10^{-17}$. It can be seen that the frequency instability of HM has improved by nearly two orders of magnitude with the steering from the optical clock. This is the best frequency instability of HMs to date, which implies that the timekeeping accuracy of the steering system based on this deployable optical clock is expected to reach a level of 100 ps per month.

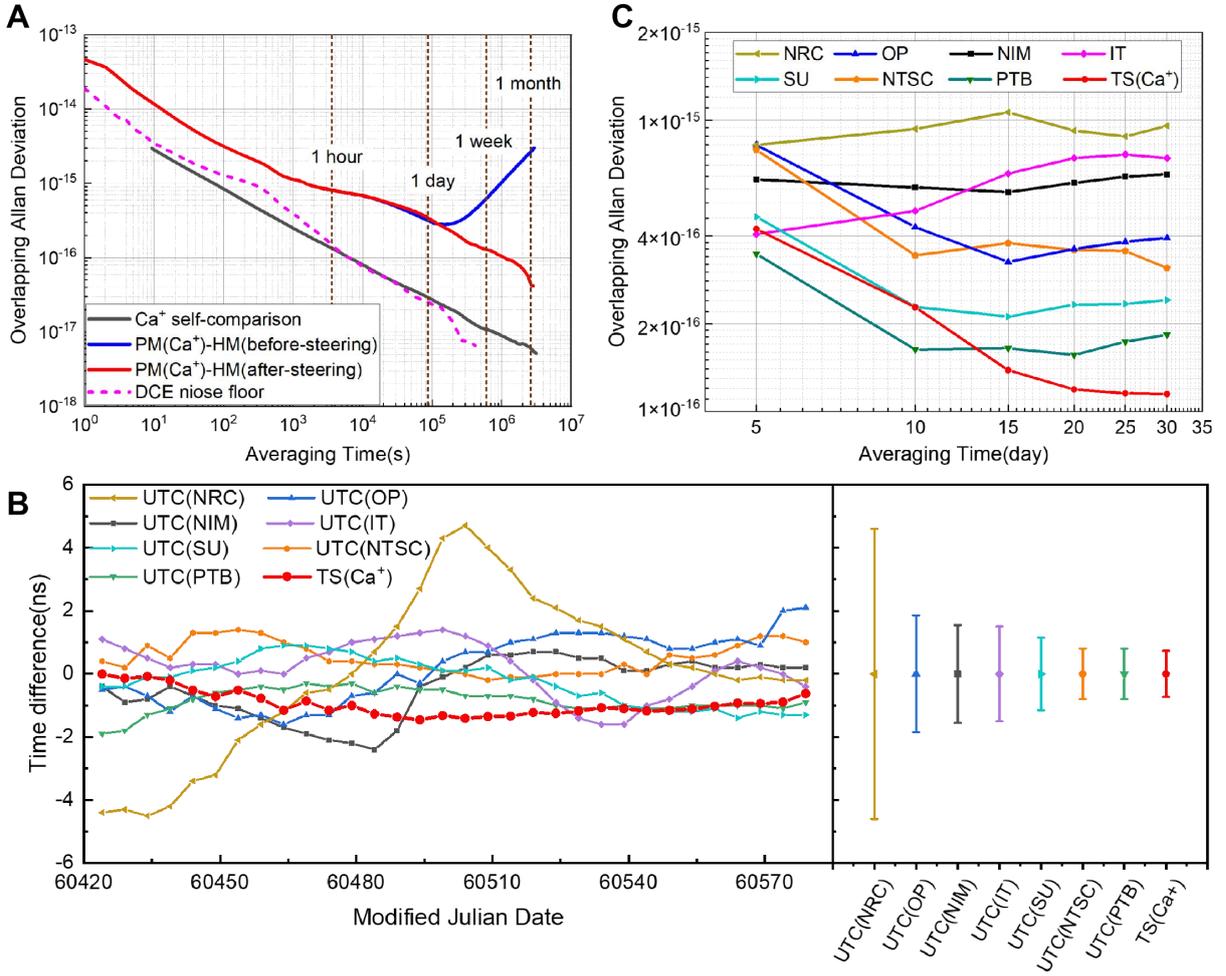

**Figure 4. Evaluation the performance of the optical time scale.** (**A**) Frequency instabilities obtained from different comparison experiments. Black solid line: self-comparison with $^{40}$Ca$^+$ optical clock; Blue solid line: comparison between the PM and the HM before steering; Red solid line: comparison between the PM and the HM after steering; Purple dotted line: contribution from the DCE noise. (**B**) Left: the time differences of UTC-TS(Ca$^+$) and UTC-UTC(k); Right: the time difference fluctuations of UTC-TS(Ca$^+$) and UTC-UTC(k). (**C**) The fractional instabilities of time difference between the UTC and UTC(k) or TS(Ca$^+$) during the period from MJD 60424 to 60579.

$$\Delta f = f_{PM} - f_{HM} \quad (1)$$

$$f_{steer} \propto \sum_{i}^{n} \Delta f_i / n \quad (2)$$

To further test the performance of this steering system, we attempted to evaluate TS(Ca$^+$) with UTC as the reference, and UTC(k) generated by other timekeeping institutions also participated in this comparison [33]. It should be noted that the time range of the sample only covers the period from MJD 60424 to 60579. On the one hand, this allows for consistency with the time points mentioned in Circular T; on the other hand, during this period, the Cs fountain clocks of all these institutions were involved in Circular T. The aggregated time difference data are shown in Figure. 4B, and all the results were calculated in the form of UTC-UTC(k). The peak-to-peak values of the time difference fluctuations of each institution are as follows: NRC: 9.2 ns; OP: 3.7 ns; NIM: 3.1 ns; IT: 3.0 ns; SU: 2.3 ns; NTSC: 1.6 ns; PTB: 1.6 ns. It is worth noting that all these values are greater than the corresponding result of TS(Ca$^+$), which has been measured precisely as 1.5 ns.

Secondly, instabilities of the timekeeping results are illustrated in Figure. 4C. When the averaging time is between 5 and 10 days, the instability of TS(Ca$^+$) appears to be worse than that of UTC(PTB), which is due to the phase noise of the satellite link. In other words, since TS(Ca$^+$) is evaluated through the TS(Ca$^+$) → TS(BSNC) → UTC(PTB) → UTC link, its link noise is greater than that of the UTC(PTB)→UTC link. However, the influence of link noise gradually decreases when the averaging time exceeds 10 days [34], and the stability of TS(Ca$^+$) gradually emerges until it surpasses UTC(PTB) when the averaging time is greater than 15 days. When the averaging time reaches 30 days, the instability of TS(Ca$^+$) attains a level of $1.15\times10^{-16}$, which is almost at the performance limit of the UTC. However, considering that the result of frequency comparison between the HM output after steering and the PM reaches $4.2\times10^{-17}$, it is reasonable to believe that the actual stability of TS(Ca$^+$) may have exceeded the limit of the UTC. Therefore, this timekeeping system based on the deployable optical clock has the ability to be employed as an ultra-stable time reference and reduce the reliance on the UTC.

## 5. Conclusion

This paper provides a brief introduction to the transportable $^{40}$Ca$^+$ optical clock (TOC-729-3) developed in our laboratory and its application in optical time scale. This compact compact clock achieves a systematic uncertainty of $1.1\times10^{-17}$ and a frequency instability of $7.5\times10^{-15}/\sqrt{\tau}$. After being transported over 1200 km, it can still be rapidly deployed and applied, demonstrating excellent reproducibility and robustness during the experiment. Thanks to advancements in automation, the clock can maintain an operational uptime of 93.6% in six months, even when operated by non-specialists. The monthly instability of the local HM has improved from $3.0\times10^{-15}$ to $4.2\times10^{-17}$ after being steered by the optical clock. It indicates that the time deviation of the optical time scale TS(Ca$^+$) from UTC is only 625 ps at the end of steering experiment, with a peak-to-peak fluctuation of merely 1.5 ns, demonstrating a substantial enhancement in the accuracy of the local time standard.

In the future, through further optimize of the steering strategy, this system will continue operating to improve the performance of the ground station time scale for the GNSS. It is particularly noteworthy that, given the widespread adoption of high-performance HMs by many international timekeeping institutions, each institution requires deploying only a limited number of optical clocks to discipline them, enabling rapid, cost-effective, and highly efficient system upgrades to achieve a local time scale with an accuracy of several hundred picoseconds per month. This is also of great significance for improving the accuracy of TAI. Additionally, for special application scenarios lacking TAI reference, establishing a compact, mobile, and high-precision time scale based on the architecture of steering HMs (or other comparable flywheels [10,35–37]) using deployable optical clocks is highly attractive, with its diverse application scenarios urgently need to be developed by us.

## Acknowledgements

This work is supported by the National Natural Science Foundation of China (Grant No.U21A20431), the Basic Frontier Science Research Program of Chinese Academy of Sciences (Grant No. ZDBS-LY-DQC028) and the Science and Technology Department of Hubei Province (Grant No. 2025AFA004).

Y.Yuan, J. Cao and J. Yuan developed the optical clock and analyzed the frequency data. D. Wang wrote the data analysis software for the transportable optical clock and the optical frequency comb. P. Fang and Q. Chen designed the down conversion electronics. S. Cao performed maintenance on the optical frequency comb. X. Wang, S. Chao and H. Shu supported the development of optical components and vacuum system. Y. Yuan, J. Yuan, D. Wang, G. Fu, Y. Yang and R. Zhao operated the steering system. G. Li, J. Xu and F. Shi analyzed the time scale data. X. Huang is the program principal investigator and J. Cao is the co-investigator. Y. Yuan, J. Cao and X. Huang prepared the manuscript. All authors read and approved the final manuscript.

# Supplementary material：

# First GNSS-deployed optical clock for local time scale upgrade

**Systematic uncertainty of the optical clock**

    The systematic uncertainty of the optical clock were evaluated both before and after being transported, and the relevant results are shown in the Table S1. TOC-729-3 eliminates the first-order Zeeman shift, quadrupole shift, and tensor Stark shift by averaging three pairs of Zeeman transitions to an extremely high level naturally. Furthermore, since the ion trap operates with the magic radio frequency [1], the shift caused by micromotion can also be ignored. The greatest contribution to the uncertainty budget comes from the blackbody radiation (BBR) frequency shift, which is comprehensively evaluated through the detailed characterization of the effective solid angle of the trapping-ion system and the monitoring of the ambient temperature during the operation of the clock [2]. After precise evaluation of each frequency shift component, the overall systematic uncertainty can always be maintained at an extremely low level of $10^{-17}$, which shows its good reproducibility. Here, we separately present the assessment of the frequency shift terms that change relatively significantly before and after the transportation, and the results are shown in Figure. S1.

**Table S1: The systematic error budgets of TOC-729-3**

| Sources | In Wuhan | | In Beijing (First week of steering) | | In Beijing (Last week of steering) | |
|---|---|---|---|---|---|---|
| | Shift/mHz | Uncertainty/mHz | Shift/mHz | Uncertainty/mHz | Shift/mHz | Uncertainty/mHz |
| Blackbody radiation AC-Stark | 358.9 | 3.9 | 360.8 | 3.9 | 360.0 | 3.8 |
| Electric quadrupole | 0.0 | 6.1 | 0.0 | 1.8 | 0.0 | 2.1 |
| Bias magnetic linear-Zeeman | 0.1 | 0.2 | 0.1 | 1.2 | 0.0 | 0.1 |
| Servo | 1.7 | 4.7 | 0.2 | 0.5 | 3.1 | 1.4 |
| Thermal motion second-Doppler | -4.8 | 0.9 | -4.8 | 0.9 | -4.8 | 0.9 |
| Excess micromotion | 0.0 | 0.4 | 0.0 | 0.4 | 0.0 | 0.4 |
| Bias magnetic second-Zeeman | 0.1 | <0.1 | 0.1 | <0.1 | 0.1 | <0.1 |
| 729 nm laser AC-Stark | 0.2 | 0.2 | 0.2 | 0.2 | 0.2 | 0.2 |
| Collision | 0.6 | 0.6 | 0.6 | 0.6 | 0.6 | 0.6 |
| Others | 0.0 | <0.1 | 0.0 | <0.1 | 0.0 | <0.1 |
| **Total** | **356.8** | **8.7** | **357.2** | **4.7** | **359.2** | **4.7** |
| **Relative/$10^{-17}$** | **86.8** | **2.1** | **86.9** | **1.2** | **87.4** | **1.2** |

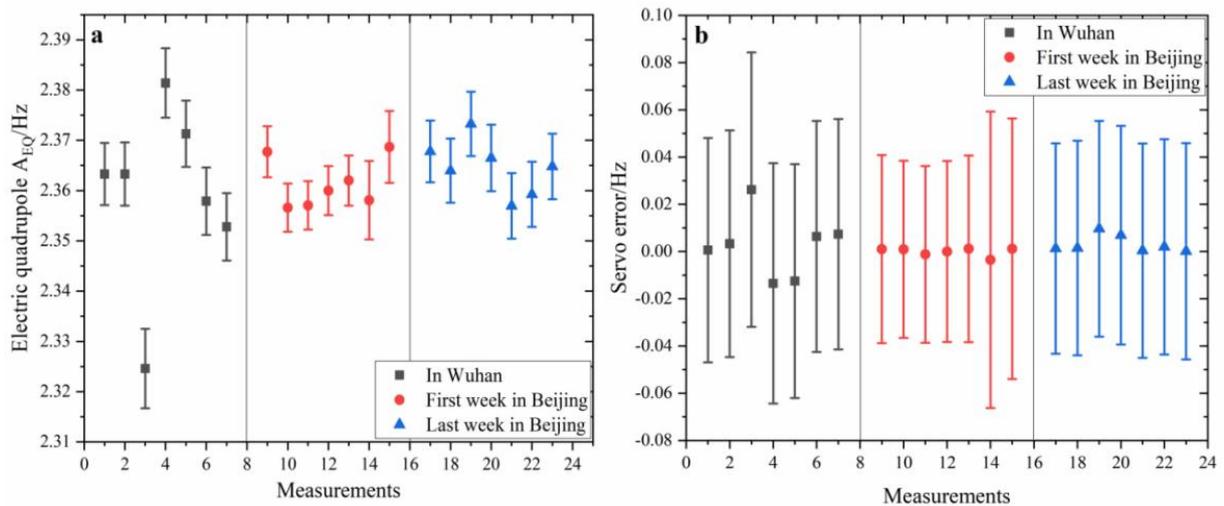

**Figure S1**. Evaluation of frequency shifts with relatively significant variation before and after the transportation of TOC-729-3. The black data points

represent the evaluation results in Wuhan, while the red and blue data points show the results from the first week and last week of steering experiments after being transported to Beijing. **a**. Evaluation results of the $A_{EQ}$ coefficient for the electric quadrupole shift [3]. **b**. Evaluation results of servo error [4].

## Down conversion electronics

The frequency down conversion of the optical clock to 10 MHz is realized by utilizing the OFC as a transfer oscillator. The principle of the down conversion electronics (DCE) is to convert the received microwave signals through a series of operations, including frequency combing, division, and differential processing, ultimately into a 10 MHz signal. When beating with the OFC, the frequency information of the clock laser turns to microwave signals, including the beat frequency $f_{beat}$, the carrier-envelope offset frequency $f_{ceo}$ and the repetition frequency $f_{rep}$:

$$f_{Ca^+} = mf_{rep} + f_{ceo} + f_{beat} \tag{S1}$$

Using these signals as inputs, the DCE can generate a 10 MHz division signal of $f_{Ca^+}$ through a series of operations.

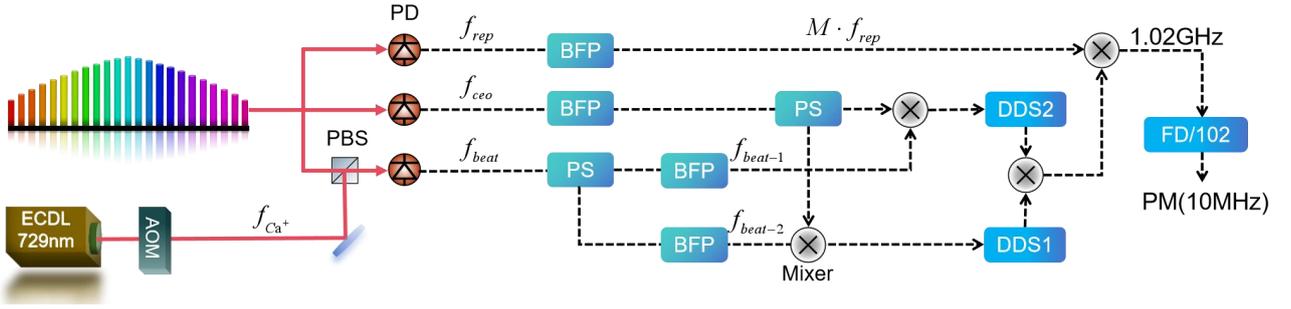

**Figure S2.** Schematic diagram of PM generation via DCE. ECDL: external-cavity diode laser; AOM: acousto-optic modulator; PBS: polarizing beam splitter; PD: photodetector; BFP: band pass filter; PS: power splitter; Mixer: double-balanced mixer; DDS: direct digital synthesizers; FD: frequency divider.

Firstly, the clock laser $f_{Ca^+}$ is divided to 1.02 GHz ($f_{TO}$) in the following manner:

$$\begin{aligned} f_{TO} &= kf_{Ca^+} = k(mf_{rep} + f_{ceo} + f_{beat}) \\ &= Mf_{rep} + \frac{km_2 - M}{m_2 - m_1}(f_{ceo} + f_{beat-1}) - \frac{km_1 - M}{m_2 - m_1}(f_{ceo} + f_{beat-2}) \end{aligned} \tag{S2}$$

The equation can be seen as removing the OFC noise from $f_{rep}$'s $M$th harmonic. $f_{beat-1}$ and $f_{beat-2}$ are the beatnotes of $f_{Ca^+}$ with OFC's $m_1$th and $m_2$th modes. Here the direct digital synthesizers (DDSs) with bit number N are the key components to realize programmable division of ($f_{ceo} + f_{beat}$). For a chosen value of $f_{TO}$, $k=c/2^N$ is determined, where $c$ is an integer that can be chosen to let the DDSs' frequency tuning words to be integers, so that the final division is free from additional quantization error.

Secondly, the 1.02 GHz signal is divided by 102 to 10 MHz via an frequency divider. During this process, the bit resolution $N$ of DDSs determines the deviation of the generated PM signal's frequency value from the ideal 10 MHz.

## Hydrogen maser noise model evaluation

We selected the hydrogen maser as the flywheel clock for optical clock steering experiments. Therefore, it is necessary to conduct an in-depth analysis of the hydrogen maser's frequency deviation, frequency drift, and model noise to rationally design the frequency steering method. Fig. 4A shows the stability of the hydrogen maser measured by the optical clock. The self-comparison stability of the optical clock and the circuit noise are both significantly lower than that of the hydrogen maser, so their influence can be neglected. And the frequency stability of the hydrogen maser evaluated using photonic microwave generated by DEC can measure the short-term to medium-term stability of the hydrogen maser accurately. When the averaging time is between $10 \sim 10^3$ s, the stability of the hydrogen maser is primarily limited by white frequency noise, approximately $3.3\times10^{-14}/\sqrt{\tau}$. When the averaging time is between $10^5 \sim 3\times10^5$ s, the hydrogen maser is mainly affected by flicker frequency noise and random walk noise. The flicker frequency noise component is about $2.4\times10^{-16}$, and the random walk noise component is approximately $1.4\times10^{-19}\sqrt{\tau}$. When the averaging time exceeds $3\times10^5$ s, the frequency stability of the hydrogen maser is dominated by frequency drift, with a drift rate of about $9.2\times10^{-22}\tau$.

**Steering method**

Since the frequency instability of the hydrogen maser (HM) reaches its minimum with an average time of approximately $10^5$ s, the steering period is set to one day. At around 1:00 (UTC+0) every day, we calculate the average value of the frequency difference between the HM and the photon microwave (PM) from 0:00 to 24:00 of the previous day, denoted as $f_{record}$. In the first 30 days of the steering experiment, the Kalman filter remains inactive due to the lack of sufficient accumulated data. During this period, the frequency drift rate $d_H$ of the HM is calculated using the available data, and then the daily calibration value $f_{steer}$ of the HM is derived based on $f_{record}$ and $d_H$. After the 30th day of the steering experiment, the Kalman filter is activated to smooth the previous data $f_{record}$ accumulated, resulting in the filtered data $f_{kalman}$. Thereafter, the daily calibration $f_{steer}$ for the output of HM is inferred using $f_{kalman}$ and $d_H$ over the past 30 days. This method utilizes the historical information of the HM to obtain the calibration directly, enabling it to handle situations such as data interruptions and outliers in a more intuitive way, thereby ensuring the accuracy and reliability of the steering experiment.

$$1\sim30 \text{ days}: f_{steer} = f_{record}(n-1) + d_H \Delta t_n \tag{S3}$$

$$31\sim166 \text{ days}: f_{steer} = f_{kalman}(n-1) + d_H \Delta t_n \tag{S4}$$

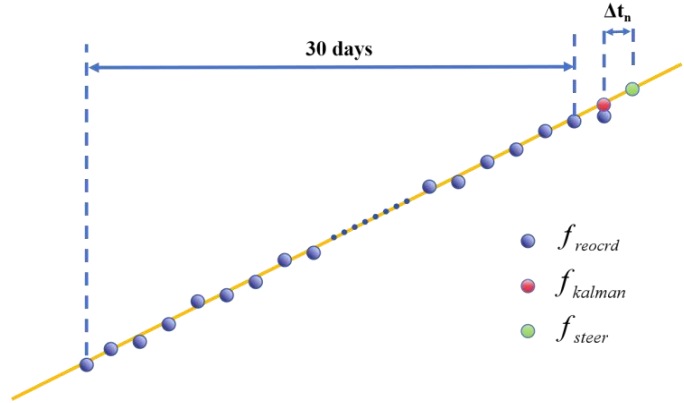

**Figure S3**. Schematic diagram of steering experiment.